\documentclass{PoS}

\title{Dynamical overlap fermions in the epsilon-regime}

\ShortTitle{Dynamical overlap fermions in the epsilon-regime}

\author{JLQCD collaboration: 
   \speaker{Hidenori~Fukaya}$^a$\thanks{E-mail: hfukaya@riken.jp},
   Shoji~Hashimoto,$^{bc}$
   Kazuyuki~Kanaya,$^d$
   Takashi~Kaneko,$^{bc}$
   Hideo~Matsufuru$^{b}$
   Kenji~Ogawa,$^e$
   Masataka~Okamoto,$^b$
   Tetsuya~Onogi,$^f$
   Norikazu~Yamada.$^{bc}$\\
   \llap{$^a$}
   Theoretical Physics Laboratory, RIKEN, Wako 351-0198, Japan\\
   \llap{$^b$}
   High Energy Accelerator Research Organization (KEK),
   Tsukuba 305-0801, Japan.\\
   \llap{$^c$}School of High Energy Accelerator Science,
             The Graduate University for Advanced Studies (Sokendai),
             Tsukuba 305-0801, Japan.\\
  \llap{$^d$} Graduate School of Pure and Applied Sciences,
           University of Tsukuba, Tsukuba 305-8571, Japan\\
  \llap{$^e$}Department of Physics, National Taiwan University,
      Taipei 10617, Taiwan.\\
  \llap{$^f$}Yukawa Institute for Theoretical Physics, Kyoto University,
             Kyoto 606-8502, Japan.
}

\abstract{
\vspace*{-90ex}
\begin{flushright}
 UTHEP-526, KEK-CP-183, RIKEN-TH-80
\end{flushright}
\vspace*{90ex}
We report on the two-flavor QCD simulation in the
$\epsilon$-regime using the overlap fermion formulation.
Sea quark mass is reduced to $\sim$
2 MeV on a $16^3\times 32$ lattice with the lattice spacing 
$a$ $\simeq$ 0.11~fm.
Topological charge is fixed at $Q=0$. 
We compare the eigenvalue distribution of the overlap-Dirac
operator with the prediction of the chiral random matrix
theory. Preliminary results on meson correlators are also
reported. 
}

\FullConference{XXIVth International Symposium on Lattice Field Theory\\
                July 23-28, 2006\\
                Tucson, Arizona, USA}

\begin{document}
\section{Introduction}

In the standard lattice QCD simulations, 
large volume, chiral, and continuum limits have to be
approached in order to obtain reliable physical quantities.
That is very challenging even with the fastest
supercomputers currently available.
In most of the previous works, the chiral limit was 
sacrificed and the simulations have been done with considerably
heavier up and down quarks, with non-chirally symmetric
Dirac operators, such as the Wilson or KS-type fermions. 

Recently, an alternative approach, which gives a priority to
the chiral limit rather than the infinite volume, is pursued
based on the understanding of the finite volume effect using
chiral effective theory. 
That is the lattice simulation in the $\epsilon$-regime.
For such a simulation, the exact chiral symmetry for the lattice
fermion is essential as one treats extremely small quark
masses of $O(\mbox{1~MeV})$.
We use the overlap Dirac operator \cite{Neuberger:1997fp}
\begin{equation}
D = m_0\left(1+\gamma_5 \mbox{sgn}H_W\right),
\end{equation}
which has the exact chiral symmetry through the 
Ginsparg-Wilson relation \cite{Ginsparg:1981bj}.
$H_W=\gamma_5D_W(-m_0)$ is the standard hermitian Wilson-Dirac operator
with large negative mass $-m_0$.

If the size of the box, $L$, is small and satisfies
\begin{equation}
  \label{eq:e-regime}
  1/\Lambda_{QCD} \ll L \ll 1/m_{\pi},
\end{equation}
where $\Lambda_{QCD}$ the QCD scale and $m_{\pi}$ the pion mass,
an expansion in terms of $\epsilon^2=m_{\pi}/\Lambda_{UV}$
($\Lambda_{UV}$ is an UV cutoff, {\it e.g.} $4\pi F_\pi$)
is valid after a careful treatment of the zero-momentum modes
\cite{Hansen:1990un,Damgaard:2001js}.
In such a small physical volume, the $\epsilon$-regime, the
low-energy constants, such as the chiral condensate $\Sigma$
and the pion decay constant $F_{\pi}$, can be extracted
through the current correlators.
Since the quark mass is already very small, the chiral 
{\it extrapolation} is not necessary.

Most previous lattice simulations in the $\epsilon$-regime
were limited to the quenched approximation \cite{Giusti:2002sm} 
except for a few pioneering works with rather heavier quarks
\cite{DeGrand:2005vb}.
The new project by the JLQCD collaboration started in March
2006 is aiming at performing large scale simulations of QCD
with the overlap fermion formulations.
We also explore the $\epsilon$-regime by pushing the quark
mass down to a few MeV.
We employ topology conserving actions to assure the overlap
fermion determinant to be smooth \cite{Fukaya:2006vs}.

For the gauge action we use the Iwasaki action with
$\beta=2.30-2.35$.
With many additional algorithmic efforts \cite{JLQCD}
we have generated thousands of configurations of two-flavor
QCD on a $16^3 \times 32$ lattice.
The lattice spacing is $a\sim 0.11-0.125$~fm.
As an exploratory run we attempted to reduce the quark mass
down to 2~MeV at $\beta=2.35$, at which $L\sim 1.8$fm.
With this small quark mass the condition (\ref{eq:e-regime})
is safely satisfied.

In this report we concentrate on the study with the smallest
sea quark mass $ma=0.002$ ($m\sim2$~MeV) at $Q=0$.
We first report on the numerical cost for this simulation 
in Sec.~\ref{sec:cost}, and describe the determination of
the lattice spacing in Sec.~\ref{sec:potential}.
Preliminary results for the low-energy constants,
$\Sigma$ and $F_{\pi}$, are obtained through the 
low-lying eigenvalues (Sec.~\ref{sec:lowmodes})
and through the pion correlators (Sec.~\ref{sec:pion}).
The summary and discussion are given in Sec.~\ref{sec:summary}.


\section{Numerical cost}
\label{sec:cost}
Naively, the simulation cost increases for small quark
masses as an inverse power of the quark mass $m$, and tends
to diverge near the massless limit.
It is, however, not the case in the finite volume.
The numerical cost or the iteration count for the solver,
say the conjugate gradient (CG) solver, is roughly
proportional to the condition number
\begin{equation}
  \label{eq:cost}
  |\lambda_{max}+m|^2/|\lambda_{min}+m|^2,
\end{equation}
where $\lambda_{min, max}$ denotes the lowest (highest)
eigenvalue of the overlap Dirac-operator.
In the large-volume limit, the numerical cost is
primarily determined by the quark mass since 
$\lambda_{min}\sim 1/\Sigma V \sim 0$
and $\lambda_{max}$ is insensitive to the gauge configuration.
In the small quark mass limit at a fixed $V$, on the other
hand, $m$ is made smaller than $1/\Sigma V$ and 
$\langle \lambda_{min}\rangle\simeq z/\Sigma V$
determines the condition number (\ref{eq:cost})
\footnote{When there are zero modes, $\lambda_{min}=0$, one must
explicitly subtract them from the determinant.}.
Here, the numerical factor $z$ can be estimated using the
chiral random matrix theory. For $N_f=2$ and $Q=0$, it is
$z\simeq 4.34$.

\begin{figure}[tbp]
  \centering
  \includegraphics[width=8cm,clip=true]{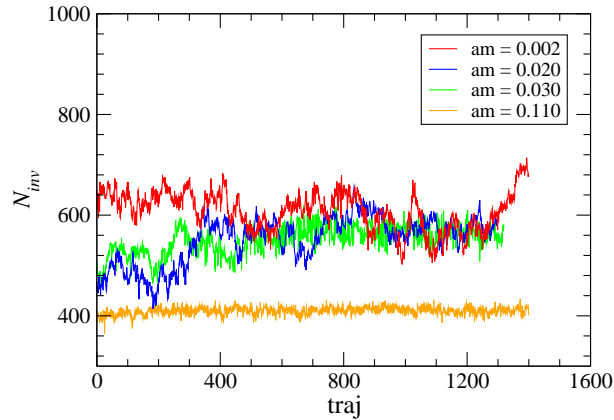}
  \caption{
 The history of CG iteration along the HMC steps.
 The sea quark mass dependence seems weak when $ma \lesssim 0.03$,
 and the numerical cost with $ma=0.002$ (red) 
 is almost the same as 
 the case with $ma=0.02$ (blue).
  }
  \label{fig:ninv}
\end{figure}

In fact, as Fig.~\ref{fig:ninv} shows, the quark mass
dependence of the iteration count is weak for 
$am\lesssim 0.03$, and the simulation cost for $am=0.002$,
which corresponds to $m\sim2$~MeV 
(with an assumption for the renormalization factor 
$Z_m=1.8$ as in the quenched theory) is almost the same as 
that for ten times heavier quark mass $am=0.020$.
However, we should note that the auto-correlation time seems
longer for smaller quark mass. 
We do not have enough statistics to precisely calculate, but
we expect it is $O(100)$ trajectories.

On a half-rack (512 nodes) of the IBM BlueGene/L (2.8~Tflops
peak performance), we need roughly one hour per trajectory.
In the following analysis, 
we use 100 configurations in $Q=0$ topological sector
sampled from 1400 trajectories 
(first 400 are discarded for the thermalization).
For the details of our numerical simulations,
we refer to the contribution by Matsufuru \cite{JLQCD}.

\section{Static quark potential with almost massless sea quarks}
\label{sec:potential}
\begin{figure}[tbp]
  \centering
  \includegraphics[width=7.4cm]{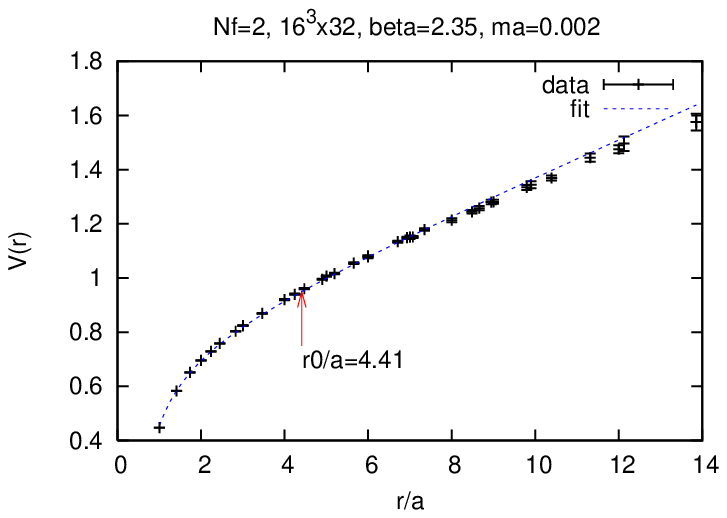}
  \includegraphics[width=7.6cm]{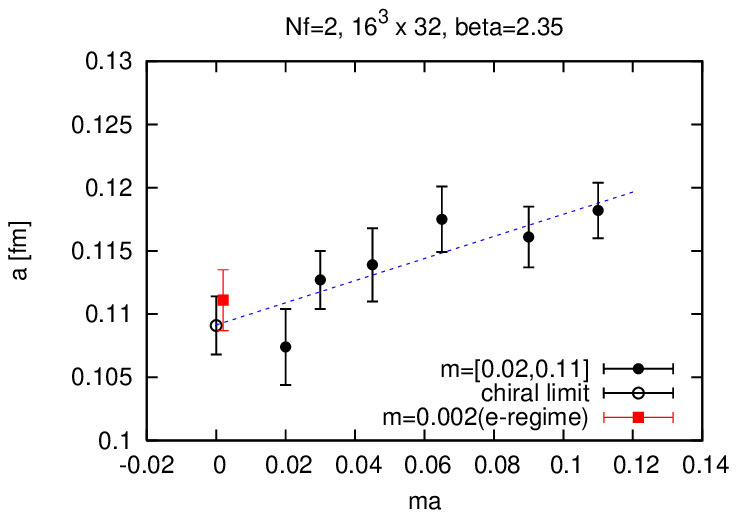}
  \caption{
  The static quark potential (left) and the lattice spacing
  with different quark mass (right).
  The quark potential has no indication of string breaking
  and $a\sim 0.11$ can be obtained, which is also 
  consistent with the chiral extrapolation from the heavier
  quark mass points, as the dotted line in the right panel shows.
  }
  \label{fig:potential}
\end{figure}

To determine the lattice spacing $a$ we measure the
static quark potential. 
In a very small quark mass regime one might expect the string
breaking, but our results in Fig.~\ref{fig:potential} (left
panel) show no indication.
This is probably because the volume is too small to contain
two static-light mesons.
The overlap of the Wilson loop with the two static-light
states could also be a problem.

We calculate the Sommer scale $r_0$ as usual and obtain
$a\sim 0.111(2)$~fm 
(with an input $r_0=$ 0.49~fm),
which is consistent with the chiral extrapolation 
from the heavier quark mass points as shown in the right panel.

\section{Low eigenmodes}
\label{sec:lowmodes}
We expect that the eigenvalue distribution of the Dirac
operator is well described by the chiral random matrix
theory (ChRMT) in the $\epsilon$-regime.
We compare them with our lattice data.

The eigenvalues of the overlap-Dirac operator lie on a
circle in the complex plane, and we project them as
\begin{equation}
  \lambda_i \equiv \frac{\mbox{Im}\lambda^{ov}_{i}}
  {1-\mbox{Re}\lambda^{ov}_{i}/2},  
\end{equation}
where $\lambda^{ov}_i$  is the $i$-th complex eigenvalues 
of the overlap-Dirac operator $D$. 
Note that $\lambda_i$ is very close to Im$\lambda^{ov}_i$
for the low-lying modes. 

\begin{figure}[tbp]
  \centering
  \includegraphics[width=7.4cm,clip=true]{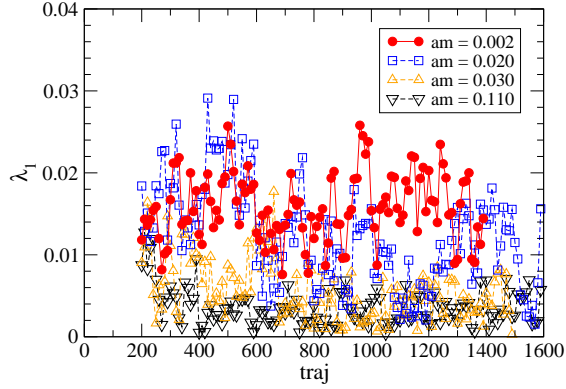}
  \caption{
    The HMC history of the lowest eigenvalue of $D$
    calculated every 10 trajectories.
  }
  \label{fig:lowhis}
\end{figure}

ChRMT predicts that the lowest mode with an eigenvalue
$\lambda_1$ in $N_f=2$ theory feels strong repulsive force
from zero. 
That is more prominent than the quenched case, and
numerically one obtains \cite{Damgaard:2000ah}
\begin{eqnarray}
\langle \lambda_1\rangle \Sigma V = 1.77 &(N_f=0),\;\;\;
\langle \lambda_1\rangle \Sigma V = 4.34 &(N_f=2),
\end{eqnarray}
where the chiral condensate $\Sigma$ could depend on the
number of flavors.
The HMC history of $\lambda_1$ shown in
Fig.~\ref{fig:lowhis} is consistent with this theoretical
expectation. 
The lowest eigenvalue is pushed up more for lighter quarks
than for heavier quarks.

\begin{figure*}[tbp]
  \centering
  \includegraphics[width=7.3cm,clip=true]{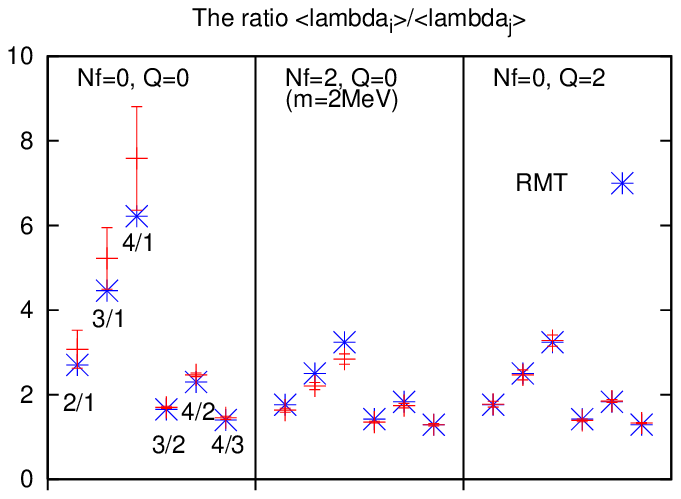}
 \hspace{0.3cm}
  \includegraphics[width=6.5cm,clip=true]{lowmode_ratio.eps}
  \caption{
    Ratio of the low-lying eigenvalues. 
    ``$i/j$'' denotes $\lambda_i/\lambda_j$.
    Left: we plot the data with $(N_f, Q)=(0,0)$, $(2,0)$, $(0,2)$
    (red crosses).
    The ChRMT predictions are shown by blue stars.
    Right: quark mass dependence of the ratios
    ``2/1'', ``3/1'' and ``4/1''.
    Crosses are ChRMT results with $(N_f, Q)=(2,0)=(0,2)$ on
    the left, and $(N_f, Q)=(0,0)$ on the right, where we
    also plot the corresponding quenched results (squares).
    Circles are lattice data in two-flavor QCD.
 }
  \label{fig:eigenratio}
\end{figure*}

Since the value of chiral condensate cannot be directly
compared for different sea quark masses, it is convenient to
look at the dimensionless ratio of the eigenvalues
$\lambda_i/\lambda_j$.
We present some of such ratios of low-lying eigenvalues on
the left panel of Fig.~\ref{fig:eigenratio}.
In the plot, the results are shown for both quenched and
unquenched ($N_f=2$) cases at $Q=0$, as well as for the
quenched data at $Q=2$.
All the data show a good agreement with ChRMT (blue star
symbols). 
ChRMT predicts that the $(N_f,Q)=(2,0)$ theory has the same
low-lying eigenvalue spectrum as the $(N_f,Q)=(0,2)$ theory,
which is well reproduced by the lattice data.

It is also interesting to see the quark mass dependence of
the ratios (the right panel of Fig.~\ref{fig:eigenratio}).
The data with $am \ge 0.03$ are close to the quenched ChRMT, 
which is shown by the crosses on the right end, and below
$am = 0.03$ they suddenly drop to the $N_f=2$ results 
(crosses on the left end).

We obtain the lattice bare value of 
$(\Sigma^{N_f=2})^{1/3}$ = 228.9(3.6)~MeV from $\lambda_1$ 
measured in $am=0.002$ configurations. 
Here, the renormalization of $\Sigma$ is not taken into
account. 

\section{Pion correlators in the $\epsilon$-regime}
\label{sec:pion}
The pion correlators in the $\epsilon$-regime
are largely deformed by the finite volume effects, 
and no longer an exponential but a quadratic function of
time $t$, 
\begin{eqnarray}
\int d^3x \langle P^a(x,t)P^b(0,0)\rangle^Q = \delta^{a,b}
[A(t/T-1/2)^2+B],\nonumber\\
A=\frac{T\Sigma^2}{F^2_{\pi}}\left[\left(
\frac{\Sigma_Q(\mu)}{\Sigma}\right)^2
+\frac{\Sigma^{\prime}(\mu)}{2\Sigma}
+\frac{\Sigma_Q(\mu)}{\mu\Sigma}\right],
B=\frac{\Sigma_Q(\mu^{\prime})}{mT},\\
\end{eqnarray}
where $\mu=m\Sigma V$ and $\mu^{\prime}=\mu(1+3\beta_1/2F^2_{\pi}V^{1/2})$.
The effective chiral condensate $\Sigma_Q(\mu)$ has strong
dependences on $Q$ and $m$, which are expressed by the
modified Bessel functions  (see \cite{Damgaard:2001js} for
the details). 
Here, $\langle O \rangle^Q$ denotes the expectation value
in the $Q$ topological sector.

\begin{figure}[tbp]
  \centering
  \includegraphics[width=7.5cm,clip=true]{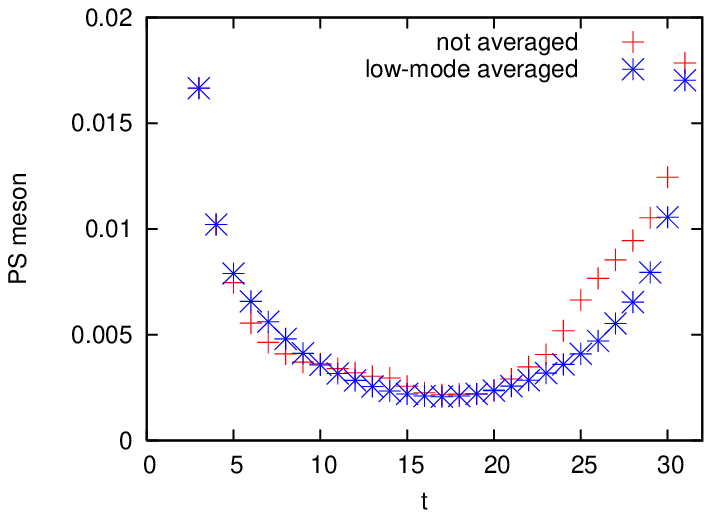}
  \includegraphics[width=7.5cm,clip=true]{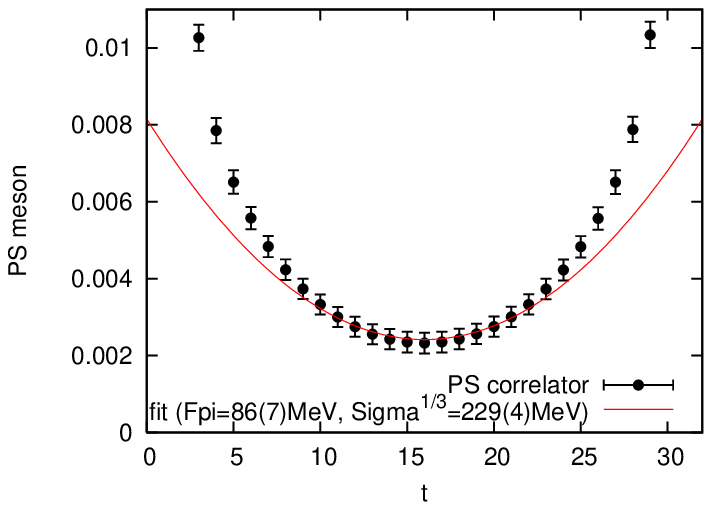}
  \caption{
    The left panel shows the effect of low-mode averaging
    (LMA) on the pion propagator(for one configuration):
    data with (blue stars) and without (red crosses) LMA are
    shown for comparison.
    A quadratic fit is attempted for the LMA pion propagator
    on the right panel.
}
  \label{fig:pion}
\end{figure}

We calculate the pion correlators on our lattice using the
low-mode averaging (LMA) technique \cite{DeGrand:2002gm}.
We find that $\sim$90\% contribution comes from the lowest 
100 modes of the Dirac operator, and the LMA is very
effective, as the left panel of Fig.~\ref{fig:pion} shows. 

With an input $\Sigma^{1/3}=229(4)$~MeV from $\lambda_1$,
the quadratic fit to the correlator works well as shown on
the right panel. 
We obtain a preliminary result $F_{\pi}=86(7)$~MeV with a
fitting range $t=[10,22]$, for which $\chi^2/\mbox{dof}\simeq 0.25$.
The result is very preliminary, not only because the
statistics is limited so far but detailed analysis of the
systematic errors is yet to be performed.

\section{Summary}
\label{sec:summary}
We pushed the dynamical overlap fermion simulation towards
the chiral limit and in fact succeeded a simulation at 
$m\sim 2$~MeV 
on the lattice with $L\sim 1.8$~fm and $a\sim 0.11$~fm.
On the finite volume lattice, the CG count does not diverge
in the chiral limit as expected from the chiral effective
theory.
The distribution of low-lying eigenvalues of $D$ is
consistent with the ChRMT expectation, from which we may
extract the value of the chiral condensate.

The measurement can be extended to the hadron correlators.
As a first test, we calculate the pion correlator and find
that the LMA works extremely well in the $\epsilon$-regime.
Precise determination of $F_\pi$ is feasible with such
calculations. 

So far, our simulation has been done at a single parameter
set, $Q=0$ and $am=0.002$.
To study other topological sectors and different quark
masses is very important, because in the $\epsilon$-regime
$Q$ and $m$ dependences are prominent and help to improve
the sensitivity to the low-energy constants.\\

Numerical studies are performed on Hitachi SR11000 and
IBM Blue Gene at High Energy Accelerator Research
Organization (KEK) under a support of its Large Scale
Simulation Program (No. 06-13).
Some parts of numerical analysis were carried out on SX8 at YITP 
in Kyoto University.
This work is supported in part by the Grant-in-Aid of the
Ministry of Education 
(No. 13135213, 16740156, 17340066, 17740171, 18034011,
18340075, 18740167).

\end{document}